\documentstyle[amssymb,pra,aps,tighten]{revtex}

\begin{document}
\title{Experimental preparation of Werner state via spontaneous parametric
down-conversion}
\author{Yong-Sheng Zhang\thanks{%
Electronic address: yshzhang@ustc.edu.cn}, Yun-Feng Huang, Chuan-Feng Li,
and Guang-Can Guo\thanks{%
Electronic address: gcguo@ustc.edu.cn}}
\address{Laboratory of Quantum Information, University of Science and Technology of\\
China, CAS, Hefei 230026, People's Republic of China\vspace{0.5in}}
\maketitle

\begin{abstract}
We present an experiment of preparing Werner state via spontaneous
parametric down-conversion and controlled decoherence of photons in this
paper. In this experiment two independent BBO (beta-barium borate) crystals
are used to produce down-conversion light beams, which are mixed to prepare
Werner state.

PACS number(s): 03.65.Ud, 42.50.-p, 03.67.-a
\end{abstract}

\section{Introduction}

Entanglement is not only one of the most striking features of quantum
mechanics, it also plays a crucial role in the field of quantum information 
\cite{Shor94,Ben84,Ekt91,Ben93}. Entangled states are important resources
for most applications of quantum information such as quantum key
distribution \cite{Ekt91}, superdense coding \cite{Ben922}, quantum
teleportation \cite{Ben93} and quantum error correction \cite{QECC} etc.
Although the best performance of such tasks requires maximally entangled
states (Bell states), the decoherence effects due to the environment make
the pure entangled state into a statistical mixture and degrade quantum
entanglement in the real word. For a practical purpose, a purification
scheme may be applied to the degraded entanglement \cite{Ben96,Deu96}.

One of the most important degraded Bell states is Werner state \cite
{Werner89}, A Werner state in $2\times 2$ system takes the following form 
\cite{Ben96}: 
\begin{equation}
\rho _W=\frac{1-F}3I_4+\frac{4F-1}3\left| \Psi ^{-}\right\rangle
\left\langle \Psi ^{-}\right| .  \eqnum{1}
\end{equation}
where $I_n$ denotes the $n\times n$ identity matrix and $\left| \Psi
^{-}\right\rangle $ is the singlet state of the four Bell states 
\begin{eqnarray*}
\left| \Phi ^{\pm }\right\rangle &=&\frac 1{\sqrt{2}}\left( \left|
00\right\rangle \pm \left| 11\right\rangle \right) , \\
\left| \Psi ^{\pm }\right\rangle &=&\frac 1{\sqrt{2}}\left( \left|
01\right\rangle \pm \left| 10\right\rangle \right) .
\end{eqnarray*}
The Werner state $\rho _W$ is characterized by a single real parameter $F$
called fidelity. This quantity measures the overlap of Werner state with a
Bell state. In the case where $F\leqslant 1/2$, the state is separable, and
thus has no entanglement to recover or maintain. And a Werner state with $%
F>\left( 2+3\sqrt{2}\right) /8\approx 0.78$ violates the
Clauser-Horne-Shimony-Holt (CHSH) inequality \cite{CHSH69,Pop94,Aravind95}.

There are many theoretical investigations on Werner state \cite{Wer}, and
the Werner state plays an important role in entanglement purification \cite
{Ben96}, nonlocality \cite{Werner89}, entanglement measure \cite{Shor01} and
etc. However, there is no report of experimental realization of Werner state
to date. At present, the most accessible and controllable source of
entanglement is obtained from the process of spontaneous parametric
down-conversion in a nonlinear crystal. In this paper, we present an
experimental preparation of bi-photon Werner state via spontaneous
parametric down-conversion and the prepared state is in the form 
\begin{equation}
\rho _W^{\prime }=x\left| \Phi ^{-}\right\rangle \left\langle \Phi
^{-}\right| +\left( 1-x\right) \frac{I_4}4,  \eqnum{2}
\end{equation}
which can be transformed to $\rho _W$ by a local unitary transformation of $%
\sigma _x\otimes I$, where $\sigma _x$ is one of Pauli operators and $I$ is
the identity operator.

There are previous works on preparing mixed single photon or bi-photon state
like in Ref. \cite{Kwiat00,White02,Thew01,ZhangC}. A. G. White {\it et al.} 
\cite{White02} have reported an optical two-qubit source that can produce a
wide range of states in the form 
\[
\rho =x\left| \Phi ^{+}\right\rangle \left\langle \Phi ^{+}\right| +\left(
1-x\right) \left| 01\right\rangle \left\langle 01\right| . 
\]
C. Zhang \cite{ZhangC} has proposed a theoretical protocol to produce an
arbitrary two-bit mixed state by using beam splitters with variable
polarization transmission coefficients and single-mode optical fibers.
However, in this paper, we propose a different way to prepare Werner states
conveniently in experiment.

This paper is organized as follows. In Sec. II, the experimental set-up is
described. The experimental results is given in Sec. III and the discussion
and summary is given in Sec. IV.

\section{Experimental Set-up}

In our experiment, the Werner state is bi-photon's polarization state, which
is obtained by mixing a complete mixed state and an entangled state.

The experimental set-up is depicted in Fig. 1. The 1.7-mm-diam pump beam at
351.1 nm (single line, 100 mW) is produced by an argon ion laser (Coherent,
Sabre, model DBW25/7), and directed to the first BBO (beta-barium borate)
crystal (cut for type-I phase matching, optic axis cut at $\theta
=35.0^{\circ }$, 1 mm thick) after passing through a polarizing beam
splitter (PBS) to give a pure horizontal polarization state. Two entangled
photon beams (polarized in vertical direction) produced by the BBO crystal
passing through a sequence of quartz plates, with an optic axis set in the
diagonal direction. It is pointed out that the photon polarization state
will decohere in such an environment \cite{Kwiat00}. The thickness of quartz
plates is set in such a way that the photon's polarization decoheres
completely (See Ref. \cite{Zhang}, the difference between optical path of
horizontal and vertical optical polarizations in the quartz plates is $%
153\lambda _0$, and $\lambda _0$ is the central wave length of the
down-converted light.). The remained pump beam passes through the BBO
crystal and is rotated to $\frac 1{\sqrt{2}}\left( \left| H\right\rangle
+\left| V\right\rangle \right) $ by a half-wave plate (HWP) and is
transmitted through the second BBO crystal which is the same as that
proposed by Kwiat in Ref. \cite{Kwiat99,White99} to produce
polarization-entangled photons in such a state. The down-converted light
beams produced by the first BBO were reflected by four reflectors and passed
through the second BBO crystal to be mixed with the down-converted beams
produced in the second BBO. Note the reason that the pump was set in single
line instead of single frequency is to decrease the coherent length (about 4
cm in our experiment) of the pump light and avoid the interference between
the down-converted light beams from the two BBO crystals.

\begin{center}
{\bf Figure 1.}
\end{center}

At the end, the bi-photon polarization state is measured tomgraphically \cite
{James01,White99} by quarter-wave plates (QWP), HWP and PBS. We use 16
analyzer settings as listed in Table I. The photons are detected by using
silicon avalanche photodiodes (EG\&G, SPCM-AQR) operated in the geiger mode.
Each detector is preceded by a small iris (the diameter is 1.5 mm) to define
the spatial mode, a narrowband interference filter (IF) centered at 702 nm
(Andover, 050FC46-25/7022 \cite{Andover}, full width at half is equal to
4.62 nm) to reduce background and define the bandwidth of the photons, and a
collection lens. The detector outputs are recorded in coincidence with a
time-to-amplitude converter and a single-channel analyzer, leading to an
effective coincidence window of 5 ns. The resulting rate of accidental
coincidences is less than $1$s$^{-1}$, which can be neglected compared with
the typical rate of the true coincidences, which is about $300$s$^{-1}$.

We also test whether the states produced by this set-up violate the CHSH
inequality. This inequality shows that $\left| S\right| \leq 2$ for any
local realistic theory, where 
\begin{equation}
S=E\left( \theta _1,\theta _2\right) +E\left( \theta _1^{\prime },\theta
_2\right) +E\left( \theta _1,\theta _2^{\prime }\right) -E\left( \theta
_1^{\prime },\theta _2^{\prime }\right)   \eqnum{3}
\end{equation}
and $E\left( \theta _1,\theta _2\right) $ is given by 
\[
\frac{C\left( \theta _1,\theta _2\right) +C\left( \theta _1^{\bot },\theta
_2^{\bot }\right) -C\left( \theta _1^{\bot },\theta _2\right) -C\left(
\theta _1,\theta _2^{\bot }\right) }{C\left( \theta _1,\theta _2\right)
+C\left( \theta _1^{\bot },\theta _2^{\bot }\right) +C\left( \theta _1^{\bot
},\theta _2\right) +C\left( \theta _1,\theta _2^{\bot }\right) }
\]
and $C\left( \theta _1,\theta _2\right) $ is the coincidence rate of two
detectors when the polarization analyzer angels are set in $\theta _1$ and $%
\theta _2$. In this experiment, we selected the settings: $\theta
_1=-22.5^{\circ }$, $\theta _1^{\bot }=67.5^{\circ }$, $\theta _1^{\prime
}=22.5^{\circ }$, $\theta _1^{\prime \bot }=112.5^{\circ }$, $\theta
_2=0^{\circ }$, $\theta _2^{\bot }=90^{\circ }$, $\theta _2^{\prime
}=45^{\circ }$, $\theta _2^{\prime \bot }=135^{\circ }$.

\section{Experimental Results}

We can adjust the intensity proportion (it can be tuned by adjusting the
position of the reflectors) between the down-converted light from two BBO
crystals to obtain Werner states with different coefficients $x$. Two output
states have been produced in this experiment, for one of them the CHSH
inequality is violated while the other is not.

The tomographic results are shown in Table I.

\begin{center}
{\bf Table I.}
\end{center}

From the data in the third column of Table I we can obtain the density
matrix of the first output state directly, however, it is not non-negative
definite \cite{James01}. We have used the maximum likelihood estimation \cite
{James01} to construct a non-negative definite density matrix 
\begin{equation}
\rho _1=\left( 
\begin{array}{cccc}
0.4169 & 0.0203+0.0022i & 0.0094-0.0237i & -0.3476+0.0296i \\ 
0.0203-0.0022i & 0.0531 & -0.0122-0.0527i & -0.0163+0.0005i \\ 
0.0094+0.0237i & -0.0122+0.0527i & 0.0559 & -0.0134-0.0191i \\ 
-0.3476-0.0296i & -0.0163-0.0005i & -0.0134+0.0191i & 0.4741
\end{array}
\right) .  \eqnum{4}
\end{equation}
The fit Werner state 
\begin{equation}
\rho _1^{\prime }=x_1\left| \Phi ^{-}\right\rangle \left\langle \Phi
^{-}\right| +\left( 1-x_1\right) \frac{I_4}4,  \eqnum{5}
\end{equation}
(where $x_1=0.801\pm 0.005$) satisfies 
\begin{eqnarray}
F\left( \rho _1,\rho _1^{\prime }\right) &=&Tr^2\left( \sqrt{\rho _1^{\prime
1/2}\rho _1\rho _1^{\prime 1/2}}\right)  \eqnum{6} \\
&=&\max F\left( \rho _1,\rho _W\right) ,  \nonumber
\end{eqnarray}
where $\rho _W$ is an arbitrary two-bit Werner state and the fidelity \cite
{Joz94} $F\left( \rho _1,\rho _1^{\prime }\right) $ is equal to $0.932$ in
this experiment. The CHSH correlation value [Eq. (3)] $\left| S\right|
=2.198\pm 0.004>2$, (the uncertainty of the HWP is $\Delta \theta \simeq
0.2^{\circ }$) which violates the CHSH inequality. The theoretical value of $%
\left| S\right| $ of state $\rho _1^{\prime }$ is equal to $2.266$.

The tomographic results of the second output are shown in the sixth column
of Table I. From these data we can obtain the density matrix of the second
output state 
\begin{equation}
\rho _2=\left( 
\begin{array}{cccc}
0.3949 & -0.0217+0.0398i & -0.0080+0.0024i & -0.1657+0.0150i \\ 
-0.0217-0.0398i & 0.1285 & 0.0165-0.0011i & -0.0075+0.0504i \\ 
-0.0080-0.0024i & 0.0165+0.0011i & 0.1311 & -0.0152+0.0228i \\ 
-0.1657-0.0150i & -0.0075-0.0504i & -0.0152-0.0228i & 0.3455
\end{array}
\right)  \eqnum{7}
\end{equation}
The fit Werner state is 
\begin{equation}
\rho _2^{\prime }=x_2\left| \Phi ^{-}\right\rangle \left\langle \Phi
^{-}\right| +\left( 1-x_2\right) \frac{I_4}4,  \eqnum{8}
\end{equation}
where $x_2=0.405\pm 0.005$ and the fidelity $F\left( \rho _2,\rho _2^{\prime
}\right) $ is equal to $0.982$. The CHSH correlation value $\left| S\right|
=1.380\pm 0.008<2$, which satisfies the CHSH inequality. The theoretical
value of $\left| S\right| $ of state $\rho _2^{\prime }$ is equal to $1.146$%
. We should note that there is significant difference in count rates between
the two sets of results. In our experiment, the intensity proportion of
down-converted light beams between two BBO crystals is tuned by adjusting
the position of the reflectors. So the length of down-converted photon's
optical path between the source and the detector has been changed for
preparing the second state. Therefore, the count rates have been decreased
significantly.

We can use the ``linear entropy'' to quantify the degree of mixture of a
quantum state and use the tangle to measure the degree of entanglement of a
two-bit quantum state \cite{James01,Wootters}. The linear entropy for a
two-bit system is defined by 
\begin{equation}
P=\frac 43\left( 1-Tr\left\{ \rho ^2\right\} \right) .  \eqnum{9}
\end{equation}
The tangle is defined as $T=C^2$, where $C$ is the concurrence in Ref. \cite
{Wootters}. In this experiment, for the first density matrix in Eq. (4), $%
P=0.46\pm 0.03$ and $T=0.35\pm 0.01$, and for the second density matrix in
Eq. (7), $P=0.83\pm 0.03$ and $T=0.01\pm 0.00$.

\section{Discussion and Conclusion}

We have prepared two typical Werner states in this experiment and measured
the density matrix tomographically. Two independent BBO crystals are used to
produce parametric down-conversion light beams which can be mixed to prepare
a wide range of two-bit quantum states. In our experiment, the intensity
proportion of down-converted light beams between two BBO crystals are tuned
by adjusting the position of the reflectors. In fact, the proportion can be
tuned with linear optical elements, which is shown in Fig. 2.

\begin{center}
{\bf Figure 2.}
\end{center}

Although the output light beams were produced by classically mixing (not
quantum superposing) the down-converted light beams, it is hard (but not in
principle) to distinguish from which crystals the output photon is produced
since the pump light is continuous. However, if we use the pulse pump we can
distinguish which crystal the output photon from by measuring the photon's
arrival time, that is, we can distill the classical information of the mixed
state and it can be regarded as a pure state.

We expect that this method can be modified to prepare arbitrary two-bit
states conveniently and this source can be used in experimental
investigation of entanglement purification or other quantum information
tasks.

\begin{center}
{\bf ACKNOWLEDGMENTS}
\end{center}

This work was supported by the National Fundamental Research Program
(2001CB309300), the National Natural Science Foundation of China, the
Innovation Funds from Chinese Academy of Sciences (CAS), and was also funded
by the outstanding Ph. D. thesis award and the CAS's talented scientist
award rewarded to Lu-Ming Duan.

\begin{center}
{\bf Appendix}
\end{center}

For a single photon, the non-dissipative coupling between photon frequency
and polarization in a birefringent media leads to decoherence of
polarization, provided we regard the freedom of frequency as ``environment''%
\cite{Kwiat00,James01,White02}.

The single-photon's initial state is described by a pure product state of
polarization and frequency, 
\begin{equation}
\left| \Psi \right\rangle =\left( a_1\left| H\right\rangle +a_2\left|
V\right\rangle \right) \otimes \int d\omega A\left( \omega \right) \left|
\omega \right\rangle .  \eqnum{A-1}
\end{equation}
with basis $\left| H\right\rangle $ (horizontal polarization) and $\left|
V\right\rangle $ (vertical polarization) denoted by 
\[
\left| H\right\rangle =\left( 
\begin{array}{l}
1 \\ 
0
\end{array}
\right) ,\left| V\right\rangle =\left( 
\begin{array}{l}
0 \\ 
1
\end{array}
\right) 
\]
respectively, and $A\left( \omega \right) $ is the complex amplitude
corresponding to $\omega $, normalized so that 
\[
\int d\omega \left| A\left( \omega \right) \right| ^2=1. 
\]
After being transmitted by the birefringent medium whose length is $x$, the
final state of the photon is 
\begin{equation}
\left| \Psi ^{\prime }\left( x\right) \right\rangle =a_1\left|
H\right\rangle \otimes \int d\omega A\left( \omega \right) e^{-\frac 12%
i\omega \left( n_H-n_V\right) x/c}\left| \omega \right\rangle +a_2\left|
V\right\rangle \otimes \int d\omega A\left( \omega \right) e^{\frac 12%
i\omega \left( n_H-n_V\right) x/c}\left| \omega \right\rangle .  \eqnum{A-2}
\end{equation}
Where $n_H$ ($n_V$) is the refractive index of horizontal (vertical)
polarization. It is necessary to note that although $n_H$ and $n_V$ are
dependent on frequency $\omega $, the value of $n_H-n_V$ does not vary
obviously with frequency varying in a small scale and we regard it as
constant. The density operator of the polarization state can be obtained by
tracing over frequency degrees of freedom from the complete density operator 
\begin{equation}
\rho \left( x\right) =\left( 
\begin{array}{ll}
a_1a_1^{*} & a_1a_2^{*}\Gamma \left( x\right) \\ 
a_1^{*}a_2\Gamma ^{*}\left( x\right) & a_2a_2^{*}
\end{array}
\right) ,  \eqnum{A-3}
\end{equation}
where 
\begin{equation}
\Gamma \left( x\right) =\int d\omega A\left( \omega \right) A^{*}\left(
\omega \right) e^{-i\omega \left( n_H-n_V\right) x/c}.  \eqnum{A-4}
\end{equation}

From Eq. (A-3) it can be seen that the evolution of the polarization's
states depend on the character of the ``environment'' which is described by
the function $A\left( \omega \right) $.

The spectrum of the photon's frequency is defined by the narrowband
interference filter (IF). In this experiment, the spectrum of the photon's
frequency is rectangular function. To obtain this environment, we use an
interference filter whose spectrum's shape is approximate to rectangle and
the full width at half maximum (FWHM) is $\delta =4.62$ nm. (Andover,
050FC46-25/7022 \cite{Andover}). The theoretical value of $\Gamma \left(
x\right) $ is 
\[
\Gamma \left( x\right) =\frac{2ic}{x\left( n_H-n_V\right) \Delta \omega }%
e^{i\omega _0x\left( n_H-n_V\right) /c}\sin \frac{x\left( n_H-n_V\right)
\Delta \omega }{2c} 
\]
and 
\begin{equation}
\left| \Gamma \left( x\right) \right| =\frac{2c}{x\left( n_H-n_V\right)
\Delta \omega }\left| \sin \frac{x\left( n_H-n_V\right) \Delta \omega }{2c}%
\right| ,  \eqnum{A-5}
\end{equation}
where we assume that $n_H>n_V.$

\begin{center}
{\bf Figure 3}.
\end{center}

In the experiment \cite{Zhang} (See Fig. 3, the BBO crystal in the figure is
cut at a degenerate type-I phase matching angle), we first measure the
density matrix of the decohered photon's polarization tomgraphically \cite
{James01}: 
\[
\rho ^{\prime }=\left( 
\begin{array}{ll}
a_1 & a_2 \\ 
a_2^{*} & a_3
\end{array}
\right) . 
\]
The experimental value of $\left| \Gamma \left( x\right) \right| $ can be
calculated as 
\begin{equation}
\left| \Gamma \left( x\right) \right| =\left| a_2/\sqrt{a_1a_3}\right| . 
\eqnum{A-6}
\end{equation}
The results are shown in Fig. 4. The coherence of the photon's polarization
degenerates as theoretical expected.. It is interesting that the coherence
is recovered to some extent after it degenerated completely, which is
similar to the Fraunhofer diffraction at a rectangular aperture (or slit) 
\cite{Born}. However, since the spectrum of the filter is not strictly
rectangular, there are some deviations at the tail of curve.

\begin{center}
{\bf Figure 4.}\bigskip 
\end{center}

Figure caption

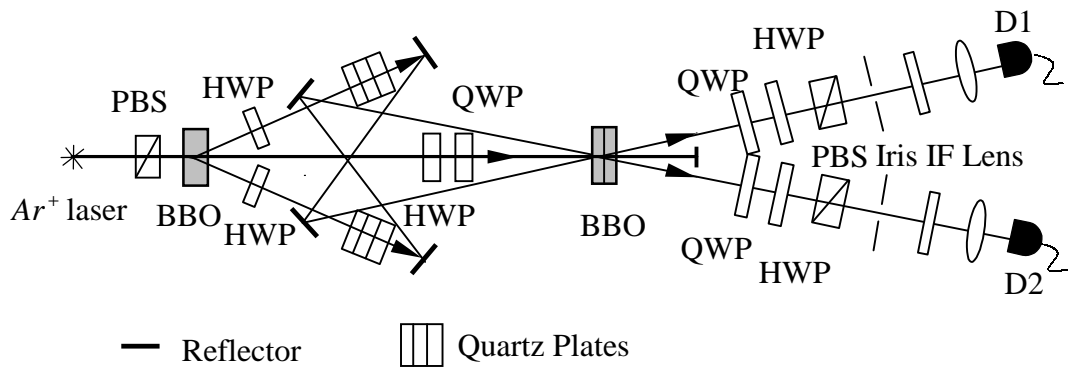
{\bf Figure 1}. Experimental set-up to produce bi-photon Werner state.

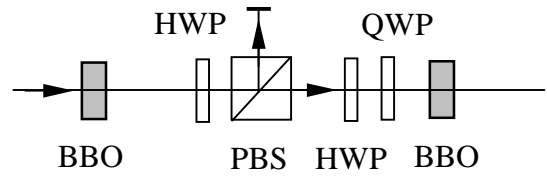
{\bf Figure 2}. Experimental set-up to tune the intensity proportion of
down-conversion light beams from two BBO crystals.

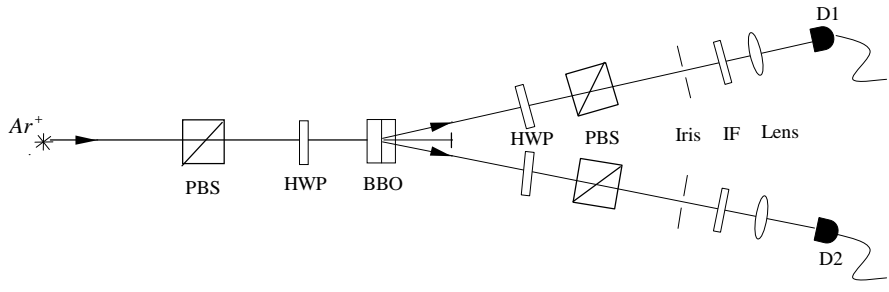
{\bf Figure 3}. Experimental setup of controlled decoherence of single
photon's polarization. The 351.1 nm line of an argon ion laser (Coherent,
Sabre) is used to pump a BBO crystal (3 mm thick), which is cut at a
degenerate type-I phase matching angle to produce a pair of entangled
photons. Each photon of the entangled pair is polarized in state $\left|
H\right\rangle $. Then the polarization of one photon of the entangled pair
is rotated to $\frac 1{\sqrt{2}}\left( \left| H\right\rangle +\left|
V\right\rangle \right) $ by the half-wave plate (HWP). After decohering in
the quartz plates, this photon's polarization state is measured
tomographically by HWP, quarter-wave plate (QWP) and polarization beam
splitter (PBS). The other photon of the pair is used to do coincidence
detection. We measure the decohered photon's polarization with 4 analysis
settings ($\left| H\right\rangle ,\left| V\right\rangle ,\left|
H\right\rangle +\left| V\right\rangle ,\left| H\right\rangle +i\left|
V\right\rangle $), allowing reconstruction of the density matrix.

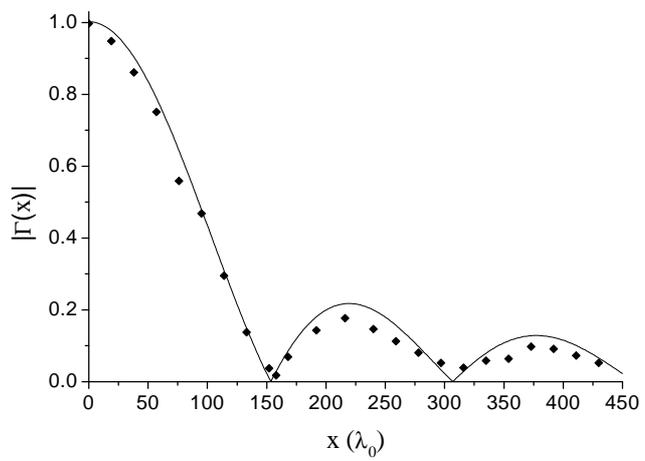
{\bf Figure 4}. Experimental results vs theoretical curve. The FWHM of the
IF is $4.62$nm and the spectrum of the frequency is assumed to be
rectangular. $x$ is the optical path difference between horizontal and
vertical polarizations and $\lambda _0$ is the central wavelength --- $702.2$%
nm --- of photons produced by SPDC.

{\bf Table I}. Settings for measuring bi-photon Stokes parameters. $H$, $V$,
and $D$ are horizontal, vertical, and diagonal ($45^{\circ }$) linear
polarization, respectively. $R$ is right circular polarization. The data are
for state $\rho _1$ (counted over 100 s).

{\bf Table II}. The data are for state $\rho _2$ (counted over 100 s).

\end{document}